\documentclass{jpp}
\usepackage{graphicx}

\usepackage[utf8]{inputenc}
\usepackage[T1]{fontenc}
\usepackage{amsmath}
\usepackage{xcolor}
\usepackage{soul}

\shorttitle{Nonlinear Quantum Simulation by Repeated Measurement}
\shortauthor{J. Andress, A. Engel, Y. Shi and S. Parker}

\title{Quantum Simulation of Nonlinear Dynamical Systems Using Repeated Measurement}

\author{Joseph Andress\aff{1}
  \corresp{\email{Joseph.Andress@colorado.edu}},
  Alexander Engel\aff{1},
  Yuan Shi\aff{1},
 \and Scott Parker\aff{1,2}}

\affiliation{\aff{1}Department of Physics, Center for Integrated Plasma Studies, University of Colorado Boulder, Boulder, CO 80309, USA\aff{2}Renewable and Sustainable Energy Institute, University of Colorado Boulder, Boulder, CO 80309, USA}

\begin{document}

\maketitle

\begin{abstract}
We present a quantum algorithm based on repeated measurement to solve initial-value problems for nonlinear ordinary differential equations (ODEs), which may be generated from partial differential equations in plasma physics. We map a dynamical system to a Hamiltonian form, where the Hamiltonian matrix is a function of dynamical variables. To advance in time, we measure expectation values from the previous time step, and evaluate the Hamiltonian function classically, which introduces stochasticity into the dynamics. We then perform standard quantum Hamiltonian simulation over a short time, using the evaluated constant Hamiltonian matrix. This approach requires evolving an ensemble of quantum states, which are consumed each step to measure required observables. We apply this approach to the classic logistic and Lorenz systems, in both integrable and chaotic regimes. Our analysis shows that solutions' accuracy is influenced by both the stochastic sampling rate and the nature of the dynamical system. 
\end{abstract}

\section{Introduction}
As modern quantum computers move closer to true utility, increasing focus has come onto the various tasks a hypothetical error-corrected quantum computer might be able to perform. The existence of an efficient quantum algorithm for solving general dynamical systems of the form
\begin{equation}
    \dot{\boldsymbol{x}} = \boldsymbol{G}(\boldsymbol{x})
    \label{eqn:GeneralSystem},
\end{equation}
is an open question which has attracted significant attention\citep{bib:LinearEmbedding,bib:Joseph_2020,bib:leyton2008quantum,bib:Liu_2021,bib:Lubasch2020,bib:Xue_2021,YuanSuggestedCubicPaper,YuanSuggestedFusionPaper}. While previous approaches, which include variational methods \citep{bib:Lubasch2020}, Carleman linearization \citep{bib:Liu_2021}, and homotopy perturbation methods \citep{bib:Xue_2021}, have shown promise in specific cases, none has yet achieved general efficiency, in the sense of a large quantum speedup with respect to the system size for general or polynomial dynamical systems without a complexity that is exponential in simulation time. For instance, the approach proposed by \citet{bib:leyton2008quantum} achieves a speedup with respect to system size for a large class of polynomial dynamical systems through a nondeterministic method based on Euler's method, but has complexity exponential in simulation time. Whereas the algorithm of \citet{bib:Liu_2021} can scale polynomially with the simulation time, but requires that the dynamical system has dissipation with strength above a certain threshold, which unfortunately excludes many systems of interest such as collisionless plasma systems.

In the special case that $\boldsymbol{G}(\boldsymbol{x})$ is linear and norm-preserving, Eq.~(\ref{eqn:GeneralSystem}) may be mapped to Schr\"odinger's equation 
\begin{equation}
    \frac{d}{dt}|\psi\rangle = -i\mathsfbi{H}|\psi\rangle
    \label{eqn:SchrodingerEquation},
\end{equation}
 and thus solutions may be extracted from established Hamiltonian simulation algorithms \citep{bib:HamSimPaper}. These algorithms are themselves not always efficient, but extensive research into Hamiltonian simulation has uncovered efficient algorithms for a number of classes of $\mathsfbi{H}$ \citep{bib:SparsePaper, bib:SparePaper2}. 
In fact, many linearized plasma problems can be converted into the Schr\"odinger form \citep{bib:AlexPaper,bib:IlyaPaper}.  

However, what makes many plasma problems challenging are nonlinearities. For example, in the Vlasov-Maxwell system, nonlinearity arises from the coupling between the particle distribution function and the electromagnetic fields. The nonlinearity of this equation, central to plasma physics, is a major hurdle in applying quantum computing to the field. The Vlasov-Maxwell system is a set of partial differential equations, but methods for ordinary differential equations would be applicable after discretizing the equation. For example, using finite difference in phase space \citep{bib:spatialdiscr}, the solution is specified only at a set of grid points, and derivatives are approximated as functions of the solution. After discretization, the remaining task is to perform time advance.

In this paper, we present a quantum algorithm for advancing nonlinear ordinary differential equations (ODEs), whose only requirements are that the dynamical system is real and the nonlinearities are polynomial. Our algorithm uses a measurement-based approach with multiple copies of the initial state, which we evolve separately and measure to determine a snapshot of the nonlinear Hamiltonian. We then use Hamiltonian simulation to advance the system forward a short period, before again measuring the system to determine a new Hamiltonian. We demonstrate our algorithm by solving the logistic equation and the widely-studied Lorentz system \citep{bib:LorenzPaper}. The algorithm presented here is relatively simple. The goal is to demonstrate how an easy-to-understand quantum algorithm can solve a real problem, building a foundation for problems of practical interest in plasma science in the future. Although not generally efficient, our measurement-based approach may efficiently solve classes of problems not covered by existing quantum algorithms. 

This paper is organized as follows. In Sec.~2, we outline the basic algorithm and present a mapping which takes a real, polynomial system to Hamiltonian form. In Sec.~3, we demonstrate the algorithm using two dynamical systems, the logistic system and the Lorenz system. In Sec.~4, we examine the algorithm's performance under the lens of quantum entropy and error. Finally, in Sec.~5, we summarize our findings. Additional details on the mapping in Sec.~2 can be found in the appendix.
 
\section{Approach}
In this paper, we attempt to generalize established methods of Hamiltonian simulation to real, polynomial systems of ODE's. Existing Hamiltonian simulation methods evolve a quantum state under a constant Hamiltonian, but by dividing the total simulation into many short times steps, we can use a different Hamiltonian for each step. Thus, the requirement of a constant Hamiltonian can be relaxed. In this paper, we generalize a constant Hamiltonian to a sum of constant Hamiltonians weighted by dynamical observables. Because the observables depend quadratically on quantum states, the Schrodinger equation has a  cubic nonlinearity. A system of ODE's can be mapped directly to cubic, nonlinear Hamiltonian form if the ODE's are cubic and norm-preserving. More generally, we present a method of mapping any real, polynomial system to cubic, norm-preserving form.

The nondeterministic nature of quantum measurements causes uncertainties in the measured quadratic observables, resulting in a stochastic spreading of the possible simulated trajectories. Using many measurements to reduce the variance, the ensemble of many such trajectories converges to the deterministic solution, and we use von Neumann entropy as a measure of this spreading.

\subsection{Piecewise linear dynamics via measurements}
For dynamical systems of the form Eq.~(\ref{eqn:GeneralSystem}), we encode elements of the vector $\boldsymbol{x}$ using amplitudes of a multi-level quantum system $\psi$:
\begin{equation}
    |x_i|^2 \propto |\langle i|\psi\rangle|^2.
    \label{eqn:AmplitudeEncoding}
\end{equation}
This amplitude encoding scheme allows $2^N$ real scalar dynamical variables to be stored on $N$ qubits. We use an approach akin to the classical forward Euler method: For the $n$-th time step, we evolve the quantum state forward by $\Delta t$ via Eq.~(\ref{eqn:SchrodingerEquation}) using a constant Hamiltonian:
\begin{equation}
    |\psi\rangle_{n} = \exp{\Big(-i\mathsfbi{H}(|\psi\rangle_{n-1})\Delta t\Big)}|\psi\rangle_{n-1}
    \label{eqn:MethodEquation}.
\end{equation}
In this unitary map, we allow $\mathsfbi{H}$ to be a function of $|\psi\rangle_{n-1}$ which we evaluate using measurements before evolving the state. Since the Hamiltonian is a constant over the time step, we can make use of linear Hamiltonian simulation algorithms. These algorithms are most efficient for sparse Hamiltonians \citep{bib:SparePaper2,bib:SparsePaper}. For more general Hamiltonians, one may use Trotter-Suzuki methods \citep{bib:Suzuki} to partition the Hamiltonian into the sum of multiple sub-Hamiltonians.

\subsection{Observable-Hamiltonian pairs for cubic nonlinear systems}
For Eq.~(\ref{eqn:MethodEquation}) to be applicable for a nonlinear dynamical system, the system must be expressible in a Hamiltonian form. The simplest nonlinear example is perhaps the cubic system, for which $\mathsfbi{H}(|\psi\rangle_{n-1})$ can be written as the sum of multiple sub-Hamiltonians, weighted by the expectation values of corresponding observables:
\begin{equation}
    \mathsfbi{H}(|\psi\rangle) = \sum_{k=1}^M \langle\psi|\mathsfbi{O}_k|\psi\rangle\mathsfbi{H}_k
    \label{eqn:ObsHamForm}.
\end{equation}
Because the observables are quadratic in $|\psi\rangle$, the dynamics $i\partial_t|\psi\rangle=\mathsfbi{H}(|\psi\rangle)|\psi\rangle$ is cubic in $|\psi\rangle$. 
For a given system, the observable-Hamiltonian pairs $\{\mathsfbi{O}_k,\mathsfbi{H}_k\}$ are constant. In other words, the pair does not depend on the quantum state, which means that they need be calculated only once, before actual evolution begins. As an example, Eq.~(\ref{eqn:LogisticObsHam}) gives the observable-Hamiltonian pair associated for the logistic system.

Because measurement collapses the state, evaluating $\mathsfbi{H}$ before each time step requires many separate copies of $|\psi\rangle_{n-1}$. By the No-Cloning theorem, a specific quantum state cannot be copied, but if the initial state and consequent evolution are stored classically, then the state can be recreated. In other words, it is impossible to directly copy $|\psi\rangle_{n-1}$ before evaluating $\mathsfbi{H}$; each copy of $|\psi\rangle_{n-1}$ must be evolved from $|\psi\rangle_0$.

Simulating to final time $T$ with time steps of $\Delta t$ requires $T/\Delta t$ separate time steps, and with $m$ measurements for each of the $M$ observables, a total of $mMT/\Delta t$ states will be consumed to evaluate $\mathsfbi{H}$ throughout the full simulation. However, it is possible to avoid storing more than one quantum state at a time, by evolving each of states separately and using classical memory to store the sampled expectation values required to recreate $\mathsfbi{H}$. This scheme is efficient if the dimension of $\mathsfbi{H}_k$ is large, but the number of observables $M$ is small. 

\subsection{Generalizing to polynomial nonlinearities: Enforcing homogeneity}
The requirement that the dynamical system be of the cubic Hamiltonian form can be relaxed, because any real system with polynomial nonlinearity can be mapped to the cubic Hamiltonian form.  
The first step of the mapping is to express the polynomial system in the tensor form
\begin{equation}
    \dot{\boldsymbol{x}} \equiv \mathsfbi{A}\boldsymbol{x}^{\otimes q}
    \label{eqn:TensorForm}.
\end{equation}
In index notation, the above equation can be written more explicitly as 
\begin{equation}
    \dot{x}_j = \sum_{\alpha} \delta_{j\alpha_1} \mathsfbi{A}_{\alpha} \prod_{k=2}^{q+1}x_{\alpha_k},
    \label{eqn:TensorExplanation}
\end{equation}
where each $\alpha$ is a multi-index containing $q+1$ indices. In Eq.~(\ref{eqn:TensorExplanation}), the Kronecker delta ensures that the only terms contributing to the derivative of $x_j$ have the index $j$ as the first entry of $\alpha$, while the remaining terms in Eq.~(\ref{eqn:TensorExplanation}) account for the coefficient $\mathsfbi{A}_{\alpha}$ and factors in $\boldsymbol{x}$ for this specific monomial term in the degree-$q$ polynomial. By this definition, all entries of $\mathsfbi{A}$ are labelled by $q+1$ indices, and any entry with first index $\alpha_1$ contributes a monomial term to the polynomial derivative of $x_{\alpha_1}$. The coefficients for these monomials are stored in $\mathsfbi{A}$, and the remaining $q$ indices correspond to the $q$ factors of $\boldsymbol{x}$ entries in the degree-$q$ monomial.

The form of Eq.~(\ref{eqn:TensorForm}) describes polynomial systems with homogeneous degree $q$. For non-homogeneous systems, we can homogenize Eq.~(\ref{eqn:GeneralSystem}) by adding a constant coordinate $x_0$ as the first component of $\boldsymbol{x}$:
\begin{equation}
        x_0 = c.
    \label{eqn:HomogeneousConstant}
\end{equation}
We multiply this constant term into lower-degree terms to raise them to degree $q$. To keep the constant term unchanged during time evolution, we add a trivial term to the dynamical system:
\begin{equation}
    \dot{x}_0 = G_0(\boldsymbol{x}) = 0.
\end{equation}
For example, in Eq.~(\ref{eqn:LogisticSystem2}), the logistic system is raised to a homogeneous degree $3$. The exact value chosen for $x_0=c$ is arbitrary, but it should be chosen to match the general scale of the other components of $\boldsymbol{x}$. In other words, it should be comparable to the other amplitudes of the normalized quantum state.

The tensor form of a system is generally not unique, as a result of the commutativity of multiplication. For example, consider the differential equation $\dot{x}_1 = 5x_1x_2x_3$. When encoding this equation into the tensor form $\dot{\boldsymbol{x}}=\mathsfbi{A}\boldsymbol{x}^{\otimes 3}$, the most general statement that can be made about $\mathsfbi{A}$ is then $\mathsfbi{A}_{1123}+\mathsfbi{A}_{1132}+\mathsfbi{A}_{1213}+\mathsfbi{A}_{1231}+\mathsfbi{A}_{1312}+\mathsfbi{A}_{1321} = 5$, because the order of factors $x_{\alpha_k}$ in Eq.~(\ref{eqn:TensorExplanation}) has no effect on the underlying dynamics.

\subsection{Rescaling to unitary evolution: Enforcement of norm-preservation}
The second step of the mapping is to rescale the dynamical system. In general, a solution to Eq.~(\ref{eqn:GeneralSystem}) cannot be encoded into a quantum state per Eq.~(\ref{eqn:AmplitudeEncoding}) without losing information on the magnitude of $\boldsymbol{x}$, because storage within a quantum state requires normalization. It is therefore necessary to find a solution to the normalized problem,
\begin{equation}
    \dot{\hat{\boldsymbol{x}}} = \boldsymbol{F}(\hat{\boldsymbol{x}}),
    \label{eqn:NormalizedSystem}
\end{equation}
where $\hat{\boldsymbol{x}}=\boldsymbol{x}/|\boldsymbol{x}|$, while also storing the overall scale of the solution. The dynamics of such a solution would be compatible with the normalization of a quantum state while allowing the exact classical trajectory to be reconstructed.

A homogeneous degree-$q$ polynomial in $\boldsymbol{x}$ scales with $a^q$ when evaluated on $a\boldsymbol{x}$. Suppose the dynamical system Eq.~(\ref{eqn:GeneralSystem}) has already been converted to the homogeneous tensor form Eq.~(\ref{eqn:TensorForm}), then
\begin{equation}
    \boldsymbol{G}(\boldsymbol{x}) = |\boldsymbol{x}|^{q} \boldsymbol{G}(\hat{\boldsymbol{x}})
    \label{eqn:HomogeneousDynamicRatio}.
\end{equation}
Taking time derivative of $\hat{\boldsymbol{x}}=\boldsymbol{x}/|\boldsymbol{x}|$, the dynamics of the normalized solution satisfy
\begin{equation}
    \dot{\hat{\boldsymbol{x}}} = |\boldsymbol{x}|^{q-1}\Big(|\hat{\boldsymbol{x}}|^2\boldsymbol{G}(\hat{\boldsymbol{x}})-[\hat{\boldsymbol{x}}\bcdot\boldsymbol{G}(\hat{\boldsymbol{x}})]\hat{\boldsymbol{x}}\Big) = |\boldsymbol{x}|^{q-1}\boldsymbol{F}(\hat{\boldsymbol{x}})
    \label{eqn:NormPreservingDynamics}.
\end{equation}
In the above equation, although $|\hat{\boldsymbol{x}}|^2=1$, we keep this term so that $\boldsymbol{F}(\hat{\boldsymbol{x}})$ is a homogeneous polynomial of degree $q+2$.
The term in the bracket is the projection operator $\delta^{ij}-\hat{x}^i\hat{x}^j$ acting on $\boldsymbol{G}$, which removes the dynamics parallel to $\boldsymbol{x}$ so that the norm of the vector is preserved. The above equation converts the original system $\dot{\boldsymbol{x}}=\boldsymbol{G}(\boldsymbol{x})$ to a norm-preserving system $\dot{\hat{\boldsymbol{x}}}=\boldsymbol{F}(\hat{\boldsymbol{x}})$. A derivation of Eq.~(\ref{eqn:NormPreservingDynamics}) can be found in Appendix \ref{appA}.
Allowing for a nonlinear mapping in time, we can write
\begin{equation}
    \begin{split}
        \frac{d}{dt'} \hat{\boldsymbol{x}} & = \boldsymbol{F}(\hat{\boldsymbol{x}}), \\
        \frac{dt}{dt'} & = |\boldsymbol{x}|^{1-q},
    \end{split}
    \label{eqn:NonlinearTime}
\end{equation}
to hide the prefactor in Eq.~(\ref{eqn:NormPreservingDynamics}). In other words, the dynamics that is parallel to $\boldsymbol{x}$ is captured as a rescaling in time. The scaling prefactor is well-defined as long as $|\boldsymbol{x}|$ is nonzero. This is guaranteed when we add the constant variable $x_0$ in Eq.~(\ref{eqn:HomogeneousConstant}), which allows for the recovery of $|\boldsymbol{x}|$ from the normalized solution, and the constant serves as a non-zero minimum on $|\boldsymbol{x}|$. In the case of the simple linear system $\dot{x}_1=\lambda x_1$, following the addition of constant $x_0$, Eq.~(\ref{eqn:NormPreservingDynamics}) yields
\begin{equation}
    \begin{split}
        \dot{\hat{x}}_0 & = -\lambda \hat{x}_0\hat{x}_1^2, \\
        \dot{\hat{x}}_1 & = +\lambda \hat{x}_0^2\hat{x}_1,
    \end{split}
\end{equation}
with $\partial_t\hat{\boldsymbol{x}}=\partial_{t'}\hat{\boldsymbol{x}}$, because $q=1$. In other words, although for the original linear system, $x_1$ depends exponentially on time, our general procedure allows it to be embedded into a norm-preserving system.

We can write the homogeneous, degree-$(q+2)$ system in Eq.~(\ref{eqn:NonlinearTime}) using a rank-$(q+3)$ tensor:
\begin{equation}
    \frac{d}{dt'}\hat{\boldsymbol{x}} \equiv \mathsfbi{F}\hat{\boldsymbol{x}}^{\otimes (q+2)}
    \label{eqn:TensorFormNP}.
\end{equation}
In other words, we introduce a tensor $\mathsfbi{F}$ corresponding to the homogeneous polynomial $\boldsymbol{F}$ such that $\boldsymbol{F}(\hat{\boldsymbol{x}})=\mathsfbi{F}\hat{\boldsymbol{x}}^{\otimes (q+2)}$. The tensor form is more easily analyzed, manipulated, and mapped to matrices, such as an observable or Hamiltonian, than the polynomial form. As an example, Eq.~(\ref{eqn:LogisticF}) gives the tensor form of the logistic system.

\subsection{Converting to Hamiltonian dynamics: Enforcement of symmetry}
Mapping the tensor $\mathsfbi{F}$ to Hamiltonian form requires that $\mathsfbi{F}$ be antisymmetric in its first two indices. The entries of $\mathsfbi{F}$ must correspond to the entires of $\mathsfbi{H}$ as defined in Eq.~(\ref{eqn:ObsHamForm}):
\begin{equation}
    \dot{\hat{\boldsymbol{x}}} = \mathsfbi{F}\hat{\boldsymbol{x}}^{\otimes(q+2)} = -i\mathsfbi{H}\hat{\boldsymbol{x}}.
\end{equation}
The Hamiltonian $\mathsfbi{H}$ is Hermitian, and the entries of $\mathsfbi{F}$ must be real for a real system. Thus, all elements of $\mathsfbi{H}$ must be pure imaginary, so that the matrix $-i\mathsfbi{H}$ is an antisymmetric, real matrix. This antisymmetry manifests in the entries of $\mathsfbi{F}$, specifically in the first index and one other index, which we choose to be the second index. Thus, we require that $\mathsfbi{F}$ be antisymmetric in its first two indices. For a general tensor, this requirement can be enforced via \citep{bib:AlexThesis}
\begin{equation}
    \mathsfbi{A}_\alpha = \frac{1}{q+3}\sum_{i=2}^{q+3}(\mathsfbi{F}_{P_{2i}\alpha}-\mathsfbi{F}_{P_{1i}P_{2i}\alpha})
    \label{eqn:FAntiSymmetry},
\end{equation}
where $\alpha$ is a multi-index of $q+3$ indices, for example, $\alpha=(1,2,3,4)$ for $q=1$. $P_{ji}$ denotes the permutation operator which swaps indices $j$ and $i$, so that $P_{23}(1,2,3,4)=(1,3,2,4)$. A derivation of Eq.~(\ref{eqn:FAntiSymmetry}) can be found in Appendix \ref{appA}. The antisymmetry of $\mathsfbi{A}$ allows the dynamics to be written in the Hamiltonian form.

As with the original tensor form, the antisymmetric tensor is not uniquely determined. Eq.~(\ref{eqn:FAntiSymmetry}) results in one of many equivalent antisymmetric tensors for a given $\mathsfbi{F}$, owing to the redundancy in tensor forms. Enforcing antisymmetry in the first two indicies reduces, but does not eliminate, this redundancy. 

The tensor yielded by Eq.~(\ref{eqn:FAntiSymmetry}) is not the only antisymmetric tensor which reproduces the desired dynamics. As with the general tensor form, the coefficient in $\mathsfbi{A}_\alpha$ may be partitioned across or combined with the entries indexed by a permutation on the latter indicies of $\alpha$, as long as the antisymmetry is preserved. Further, although we have chose $\mathsfbi{A}$ to be antisymmetric in its first two indices, in principle, a mapping to Hamiltonian form is possible as long as $\mathsfbi{A}$ is antisymmetric under the exchange of the first index and any of the latter indices. The choice of the second index is arbitrary.

\subsection{Mapping to standard form: Enforcement of cubic degree}
The standard form Eq.~(\ref{eqn:SchrodingerEquation}) and Eq.~(\ref{eqn:ObsHamForm}) together yield a cubic system, with two powers of $|\psi\rangle$ coming from the expectation of the observable in Eq.~(\ref{eqn:ObsHamForm}) and the final power included in Eq.~(\ref{eqn:SchrodingerEquation}). To map a general dynamical system to the standard form, the degree-$(q+2)$ system encoded in the rank-$(q+3)$ tensor $\mathsfbi{A}$ must be reduced to a degree-$3$ system in a rank-$4$ tensor.

For odd $q$, we can apply the degree-reduction technique \citep{bib:AlexThesis}, which evolves polynomials in $\hat{\boldsymbol{x}}$
\begin{equation}
    \hat{\boldsymbol{y}}_\alpha \propto \prod_{i=1}^{(q+1)/2} \hat{\boldsymbol{x}}_{\alpha_i}
    \label{eqn:MDegreeReduction1},
\end{equation}
using a modified tensor form
\begin{equation}
    \frac{d}{dt'}\hat{\boldsymbol{y}} = \mathsfbi{M}\hat{\boldsymbol{y}}^{\otimes 3},
    \label{eqn:MDegreeReduction2}
\end{equation}
which preserves the system's dynamics. The exact form of $\mathsfbi{M}$ is not uniquely determined, but one possible choice is
\begin{equation}
    \mathsfbi{M}_{\alpha\beta\nu\eta} = \sum_{i=1}^{(q+1)/2}\Big[\prod_{j\neq i}\delta_{\alpha_j\beta_j}\Big]\mathsfbi{A}_{\alpha_i\beta_i\nu\eta}
    \label{eqn:MDegreeReduction3}.
\end{equation}
Here, the variables $\alpha$, $\beta$, $\nu$, and $\eta$ each denotes a multi-index of $(q+1)/2$ indices. In the case of even $q$, an extra constant factor of $x_0$ can be used to raise the system to odd degree. Applying the degree-reduction technique to the antisymmetric tensor $\mathsfbi{A}$ yields a cubic system, which maps to Eq.~(\ref{eqn:ObsHamForm}). As an example, Eq.~(\ref{eqn:LogisticM}) gives the reduced tensor form of the logistic system.

\subsection{Determining Observable-Hamiltonian Pairs}
Having found the rank-4 tensor $\mathsfbi{M}_{a\beta\mu\nu}$ which is antisymmetric in its first two indices, it is possible to extract the Hermitian observables and Hamiltonians appearing in Eq.~(\ref{eqn:ObsHamForm}). The real system has two equivalent forms:
\begin{equation}
    \begin{split}
        \frac{d}{dt'}y_\alpha & = \sum_{\beta,\nu,\eta}\mathsfbi{M}_{\alpha\beta\nu\eta}y_\beta y_\nu y_\eta, \\
        \frac{d}{dt'}y_\alpha & = -i\sum_{\beta,\nu,\eta}\Big(\sum_{j,k}\mathsfbi{O}^{(\nu\eta)}_{jk}y_jy_k\Big)\mathsfbi{H}^{(\nu\eta)}_{\alpha\beta} y_\beta.
    \end{split}
    \label{eqn:EquivForm}
\end{equation}
In the later form, raised indices denote a specific observable-Hamiltonian pair, while lowered indices denote a position within a matrix.

We have ensured that the first two indices of the tensor $\mathsfbi{M}$ are antisymmetric, while the order of the latter pair of indices is arbitrary. This freedom allows us to chose $\mathsfbi{M}_{\alpha\beta\nu\eta}$ to be non-zero only if $\eta\geq\nu$. The remaining requirement is that the matrices $\mathsfbi{O}^{(\nu\eta)}$ and $\mathsfbi{H}^{(\nu\eta)}$ are Hermitian.
A viable mapping from the first to the second form in Eq.~(\ref{eqn:EquivForm}) is
\begin{equation}
    \label{eqn:OHpair}
    \begin{split}
        \mathsfbi{O}_{jk}^{(\nu\eta)} & = \frac{1}{2}\Big(\delta_{\nu j}\delta_{\eta k}+\delta_{\nu k}\delta_{\eta j}\Big), \\
        \mathsfbi{H}_{\alpha\beta}^{(\nu\eta)} & = i\mathsfbi{M}_{\alpha\beta\nu\eta}.
    \end{split}
\end{equation}
It is worth pointing out that Eq.~(\ref{eqn:OHpair}) is not necessarily the optimal mapping that produces the least number of observable-Hamiltonian pairs. For example, as we shall see in Sec.~\ref{sec:logistic}, this mapping requires two pairs for the logistic system, for which one pair would be sufficient, owing to the fact that the two pairs generated by this algorithm share a Hamiltonian, up to a constant. Although not optimal, Eq.~(\ref{eqn:OHpair}) is applicable to general systems. 

\subsection{Algorithm Scaling}
Having described how to map a general nonlinear system [Eq.~(\ref{eqn:GeneralSystem})] to the standard cubic Hamiltonian form [Eq.~(\ref{eqn:ObsHamForm})] that is amenable to quantum Hamiltonian simulation using piece-wise linearization [Eq.~(\ref{eqn:MethodEquation})], let us briefly discuss the numerical cost of our approach:
Measuring an observable's expectation value to an accuracy $\epsilon$ requires $\mathcal{O}(\epsilon^{-1})$ measurements. Classically, this scaling is $\mathcal{O}(\epsilon^{-2})$ by the Central Limit Theorem, but quantum storage allows for improved scaling under amplitude amplification techniques \citep{bib:Brassard_2002}. Measuring all $M$ observables over a simulation time $T$ thus requires $\mathcal{O}(\frac{MT}{\epsilon\Delta t})$ total measurements, and each measurement destroys one copy of the current state $|\psi\rangle_{n-1}$. On average, each state will be evolved for half the total runtime, meaning the number of Hamiltonian simulation steps is $\mathcal{O}(\frac{MT^2}{2\epsilon\Delta t^2})$. The total cost of the measurements and Hamiltonian simulation steps are then $\mathcal{O}(g\frac{MT}{\epsilon\Delta t})$ and $\mathcal{O}(h\frac{MT^2}{2\epsilon\Delta t^2})$, respectively, where $g$ is the cost of a single measurement and $h$ the cost of a single evolution step. In general, $g$ and $h$ depend on a number of factors. For example, certain observables are more easily measured than others, and Hamiltonian simulation is most efficient for sparse matrices.

The scaling of the algorithm depends on the number of observable-Hamiltonian pairs, but also the exact nature of those pairs and whether the final total Hamiltonian can be evolved forward efficiently. For most systems, this algorithm is unlikely to be efficient, but there may exist specific systems, potentially those which can be realized with a few, sparse sub-Hamiltonians, for which the algorithm has a chance of outperforming alternative methods.

\subsection{Convergence to the Stochastic Model}
Our approach uses measurements to achieve linearization, which injects statistical fluctuations into otherwise deterministic systems.  
Evaluating Eq.~(\ref{eqn:ObsHamForm}) at the beginning of each time step requires finding $\langle\psi|\mathsfbi{O}_k|\psi\rangle$. At the beginning of each step, $m$ measurements of $\mathsfbi{O}_k$ generate a sample mean for the expectation value $\langle\psi|\mathsfbi{O}_k|\psi\rangle$. As $m$ increases to infinity, this sample mean converges to the exact value, but for finite $m$, the sample is a multinomial random variable, with non-zero variance in its mean. Variation in the measured values of $\langle\psi|\mathsfbi{O}_k|\psi\rangle$ causes spreading in simulated trajectories and divergence from the exact solution. For sufficiently large $m$ and small $\Delta t$, the quantum algorithm converges to the solution of a stochastic differential equation, with the system's stochasticity quantified by the rate $s=m/\Delta t$ of measurements per unit time. As $s$ increases to infinity, either by increasing the number of measurements or shortening time steps, the quantum simulation and its stochastic model converge to the exact solution. However, over a total simulation time $T$, the quantum algorithm requires $MTs$ measurements. In this paper, we present a deterministic, $s\to\infty$ solution, as a stand-in for the exact solution, alongside many finite $s$, quantum trajectories. Finding the $s\to\infty$ solution is only possible on on classical computers.

\section{Algorithm Demonstrations}
To demonstrate how to apply our approach to general polynomial nonlinear system, we use the logistic system and the Lorenz system as two examples. The logistic system is used as a simple, easily-verifiable test case, and the Lorenz system is used as a chaotic model of dissipative fluid mechanics, and thus as a precursor to plasma mechanics.

\subsection{Solutions to the Logistic System \label{sec:logistic}}
The logistic system is commonly used to describe populations and other systems which grow to a maximum value \citep{bib:verhulst1838notice}. In normalized units, the logistic system is described by the quadratic ODE:
\begin{equation}
    \dot{x}_1 = x_1(1-x_1)
    \label{eqn:LogisticSystem}.
\end{equation}
Here, we use the logistic system as a simple demonstration of our approach. For this system, the mapping is simple enough to be completed by hand. As a first step, we introduce the constant $x_0=1$ to this system,
\begin{equation}
    \begin{split}
        \dot{x}_0 & = 0, \\
        \dot{x}_1 & = x_0^2x_1-x_0x_1^2,
    \end{split}
    \label{eqn:LogisticSystem2}
\end{equation}
to raise it to a homogeneous degree-3 system.

Using Eq.~(\ref{eqn:NormPreservingDynamics}), without the prefactor, we generate the norm-preserving system:
\begin{equation}
    \begin{split}
        \frac{d}{dt'}\hat{x}_0 & = +\hat{x}_0^2\hat{x}_1^3-\hat{x}_0^3\hat{x}_1^2 = \mathsfbi{A}_{010011}\hat{x}_1\hat{x}_0\hat{x}_0\hat{x}_1\hat{x}_1 + \mathsfbi{A}_{010001}\hat{x}_1\hat{x}_0\hat{x}_0\hat{x}_0\hat{x}_1, \\
        \frac{d}{dt'}\hat{x}_1 & = -\hat{x}_0^3\hat{x}_1^2+\hat{x}_0^4\hat{x}_1^1 = \mathsfbi{A}_{100011}\hat{x}_0\hat{x}_0\hat{x}_0\hat{x}_1\hat{x}_1 + \mathsfbi{A}_{100001}\hat{x}_0\hat{x}_0\hat{x}_0\hat{x}_0\hat{x}_1,
    \end{split}
\end{equation}
corresponding to four non-zero elements of $\mathsfbi{A}$:
\begin{equation}
    \begin{split}
        \mathsfbi{A}_{100001} = -\mathsfbi{A}_{010001} & = 1, \\
        \mathsfbi{A}_{010011} = -\mathsfbi{A}_{100011} & = 1.
    \end{split}
    \label{eqn:LogisticF}
\end{equation}

For this simple system, we can find $\mathsfbi{M}$ manually:
\begin{equation}
    \begin{split}
        \mathsfbi{M}_{(10)(00)\mu\nu} = \mathsfbi{M}_{(11)(01)\mu\nu} = \mathsfbi{M}_{(01)(00)\mu\nu} = \mathsfbi{M}_{(11)(10)\mu\nu} & = \mathsfbi{A}_{(1)(0)\mu\nu} = +1, \\
        \mathsfbi{M}_{(00)(10)\mu\nu} = \mathsfbi{M}_{(01)(11)\mu\nu} = \mathsfbi{M}_{(00)(01)\mu\nu} = \mathsfbi{M}_{(10)(11)\mu\nu} & = \mathsfbi{A}_{(0)(1)\mu\nu} = -1, \\
        \mathsfbi{M}_{(00)(10)\mu\xi} = \mathsfbi{M}_{(01)(11)\mu\xi} = \mathsfbi{M}_{(00)(01)\mu\xi} = \mathsfbi{M}_{(10)(11)\mu\xi} & = \mathsfbi{A}_{(0)(1)\mu\xi} = +1, \\
        \mathsfbi{M}_{(10)(00)\mu\xi} = \mathsfbi{M}_{(11)(01)\mu\xi} = \mathsfbi{M}_{(01)(00)\mu\xi} = \mathsfbi{M}_{(11)(10)\mu\xi} & = \mathsfbi{A}_{(1)(0)\mu\xi} = -1, \\
    \end{split}
    \label{eqn:LogisticM}
\end{equation}
where $\mu = (00)$, $\nu=(01)$, and $\xi=(11)$. In Eq.~(\ref{eqn:LogisticM}), parentheses have been added to the multi-indices of $\mathsfbi{A}$ to highlight the relationship with the multi-indices of $\mathsfbi{M}$; these parentheses have no significance other than indicating the grouping of indices. We then map this result to the observable-Hamiltonian form accepted by the algorithm:
\begin{equation}
    \mathsfbi{O} = \begin{bmatrix}
        0 & \frac{1}{2} & 0 & -\frac{1}{2} \\
        \frac{1}{2} & 0 & 0 & 0 \\
        0 & 0 & 0 & 0 \\
        -\frac{1}{2} & 0 & 0 & 0 \\
    \end{bmatrix}; \;
    \mathsfbi{H} = \begin{bmatrix}
        0 & -i & -i & 0 \\
        i & 0 & 0 & -i \\
        i & 0 & 0 & -i \\
        0 & i & i & 0 \\
    \end{bmatrix}.
    \label{eqn:LogisticObsHam}
\end{equation}

\begin{figure}
    \centering
    \includegraphics[width = 0.75\linewidth]{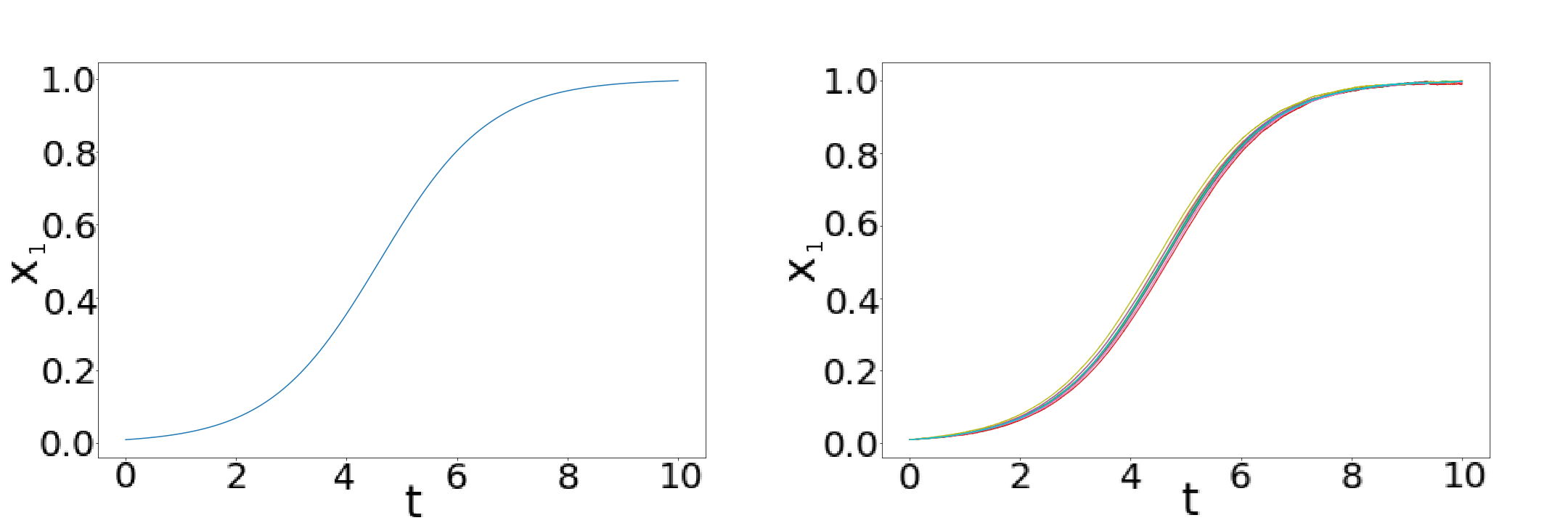}
    \caption{Time evolution of the logistic system with initial condition $x_1(0)=10^{-2}$. The deterministic trajectory (a)  can be recovered as the average of a collection of 10 quantum trajectories (b), which use the stochastic parameter $s=5\times10^5$.}
    \label{fig:LogisticSystem}
\end{figure}

After encoding the system into the quantum amplitudes and evolving as described previously, we recover the final position of the trajectory by finding the value of $\hat{x}_0$, using this to determine the final value of $|\boldsymbol{x}|$, and thus recovering $x_1$. The results of the algorithm on a classical simulator are shown Figure \ref{fig:LogisticSystem}. For simple and convergent dynamics like the logistic system, the underlying behavior dominates over any stochastic fluctuations, resulting in minimal spreading. Under these conditions, the quantum trajectories closely follow the deterministic solution.

\subsection{Solutions to the Lorenz System}
As the next nontrivial example, we apply our approach to the Lorenz system \citep{bib:LorenzPaper}:
\begin{equation}
    \begin{split}
        \dot{x}_1 & = \sigma(x_2-x_1), \\
        \dot{x}_2 & = x_1(\rho-x_3)-x_2, \\
        \dot{x}_3 & = x_1x_2-\beta x_3.
    \end{split}
    \label{eqn:LorenzSystem}
\end{equation}
Originally used to describe atmospheric flow, the Lorenz system is a classic example of chaotic dynamics, although the system is only chaotic under certain parameters \citep{bib:LorenzDetails}: $\beta \approx 8/3$, $\rho \approx 28$, $\sigma \approx 10$, and away from this region, the flow is integrable.
The system generally has multiple fixed points, at the origin and the points
\begin{equation*}
    \Big(\pm\sqrt{\beta(\rho-1)},\pm\sqrt{\beta(\rho-1)},\rho-1\Big).
\end{equation*}
Outside the chaotic regime, these fixed points are attractors, with the system's overall behavior highly dependent on the behavior around these points. The chaotic system follows a butterfly-shaped trajectory around two of these points, with the solution switching between the two unpredictably. The inherent chaos makes the Lorenz system a useful bridge to plasma physics, which may also encounter chaotic dynamics, for example, with turbulent flows.

The homogeneous Lorenz system is degree-2, but it is necessary to add an extra constant factor to raise it to degree-3 for later degree reduction, which requires odd degrees. For the cubic system, the norm-preserving system in Eq.~(\ref{eqn:NormPreservingDynamics}) is degree-5, per Eq.~(\ref{eqn:TensorFormNP}). By mapping to quadratic polynomials in the entries of $\hat{\boldsymbol{x}}$, we reduce the system to degree-3 on $16$ dynamic variables. The $16$-variable state of the system can be stored on $4$ qubits, and the final mapping requires at most $26$ observable-Hamiltonian pairs, although this may not be optimal. We automate the mapping protocol using an openly available Python code \citep{bib:GitLabRepo}.

\begin{figure}
    \centering
    \includegraphics[width = \linewidth]{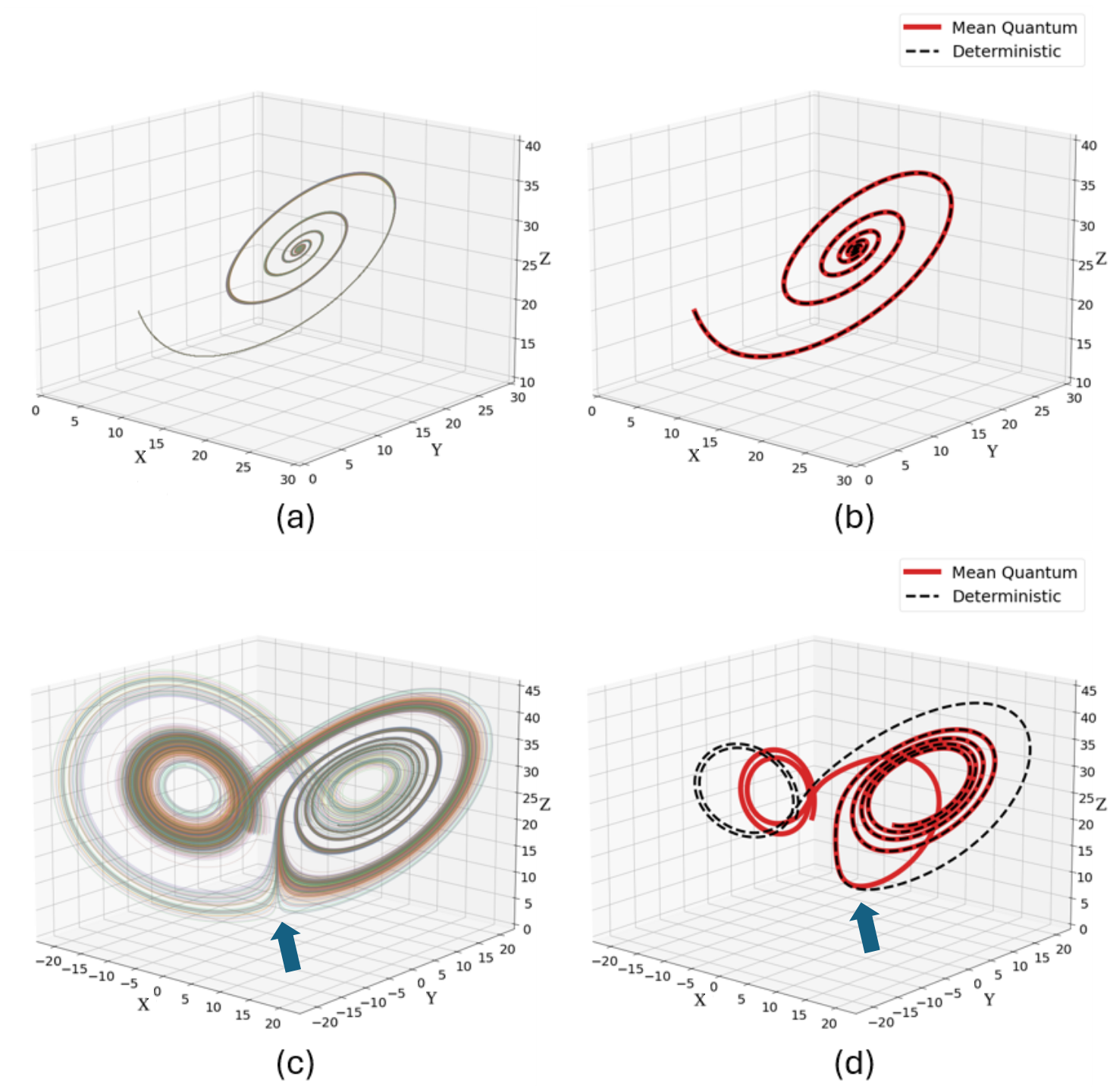}
    \caption{Time evolution of the well-behaved (a,b) and chaotic (c,d) Lorenz system with parameters $\rho=28$, $\sigma=10$, $\beta=10$ (well-behaved), $\beta=8/3$ (chaotic) and initial conditions $\boldsymbol{x}=(4.856, 7.291, 18.987)$. A collection of 300 independent, stochastic quantum trajectories (a,c) with $\Delta t=10^{-5}$, $s=10^{15}$ yield a mean trajectory (red) which initially follows the $\Delta t=10^{-5}$, $s\to\infty$, deterministic solution (black) (b,d). At the point indicated by arrows in (c,d), the chaotic trajectories diverge, resulting in eventual deviation from the deterministic solution.}
    \label{fig:ChaoticLorenz}
\end{figure}

We run our algorithm for example problems. First, for a well-behaved Lorenz system (Fig.~\ref{fig:ChaoticLorenz}a,b), the trajectories remain together for the entire simulation. By the time that noticeable spread would occur, the system has converged to its steady-state solution. Second, for a chaotic Lorenz system (Fig.~\ref{fig:ChaoticLorenz}c,d), stochastic spread in the trajectories causes significant divergence at later times. Of primary interest is the branching point indicated by the arrow. This branching point corresponds to the first sizeable divergence between the mean trajectory and deterministic solution, as shown in Fig.~\ref{fig:ChaoticLorenz}d.

\section{Error and Entropy}

We need a measure to compare a classical, deterministic solution to a collection of stochastic trajectories modelled on quantum hardware. In addition, we seek a method of characterizing the internal divergence of the trajectories with this collection.

\subsection{Quantifying stochasticity: Ensembles and entropy}
Due to its stochastic nature, the algorithm produces a different trajectory each time, even for fixed systems, initial conditions, and  sampling rates. Taking the quantum states $|\psi\rangle$ of $K$ trajectories as an ensemble, we can define a density matrix $\rho$ for the states as
\begin{equation}
    \rho = \frac{1}{K}\sum_{j=1}^K |\psi_j\rangle\langle\psi_j|.
    \label{eqn:DensityMatrix}
\end{equation}
As the trajectories spread due to the algorithm's stochasticity, the von Neumann entropy $S$ \citep{bib:NCBook} of $N$ qubits,
\begin{equation}
    S = -\text{Trace}[\rho\log{\rho}] \leq N\log{2},
\end{equation}
of this density matrix increases. At the beginning of the simulation, each of the trajectories has the exact same initial state, and thus the density matrix describes a pure state, with zero entropy. For a deterministic simulation, the states of the ensemble would remain identical, and the entropy does not increase. However, the stochasticity of the quantum simulation introduces small impurities to the ensemble, which become more pronounced as the simulation progresses. Thus, under a stochastic scheme, von Neumann entropy increases with the runtime.

In addition to measuring the intrinsic spread of an ensemble, we can also compare the quantum ensemble to a pure state, which is formed by sampling the quantum states in the deterministic trajectory on classical computers. The difference between two ensembles encoded in $\rho_1$ and $\rho_2$ can be quantified by the trace distance
\begin{equation}
    T(\rho_1,\rho_2) = \frac{1}{2}||\rho_1-\rho_2||_1.
\end{equation}
Encoding the deterministic solution as the amplitudes of a quantum state, and then as a density matrix, allows us to use the trace distance as a measure of error between the quantum ensemble and deterministic solution.

\subsection{Error and entropy of the chaotic Lorenz system}

\begin{figure}
    \centering
    \includegraphics[width = 0.8\linewidth]{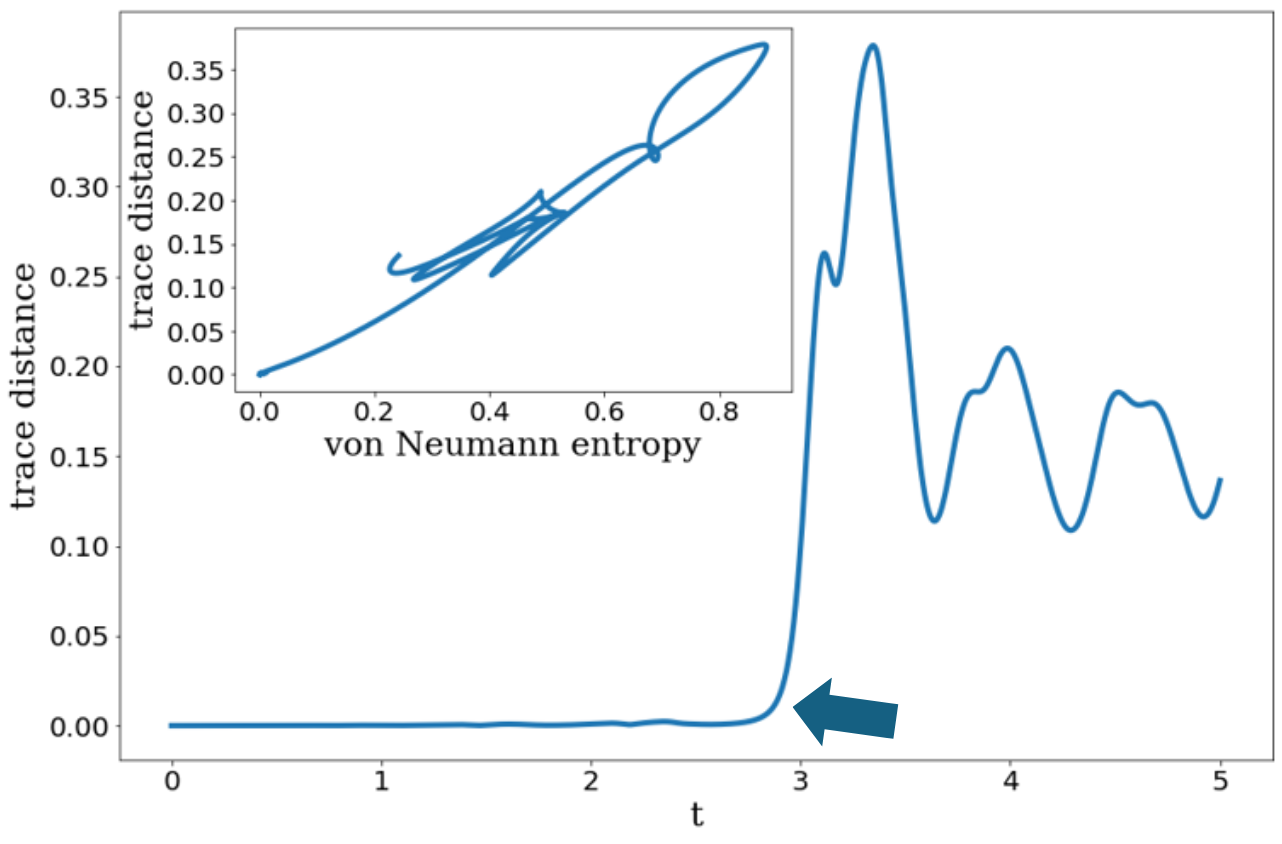}
    \caption{Trace distance between the ensemble of trajectories in Figure \ref{fig:ChaoticLorenz}c and the deterministic solution increases suddenly at the indicated branching point, and closely correlates to the von Neumann entropy (inset).}
    \label{fig:EntropyData}
\end{figure}

Taking the trajectories in Fig. \ref{fig:ChaoticLorenz}c as an ensemble, the increase in trace distance to the deterministic solution is shown in figure \ref{fig:EntropyData}. As demonstrated in the inset, the von Neumann entropy and the trace distance (error) between the ensemble and deterministic trajectories are strongly correlated. The arrow in Figure \ref{fig:EntropyData} corresponds to the branching point in Figure \ref{fig:ChaoticLorenz}c,d.

\begin{figure}
    \centering
    \includegraphics[width = 0.625\linewidth]{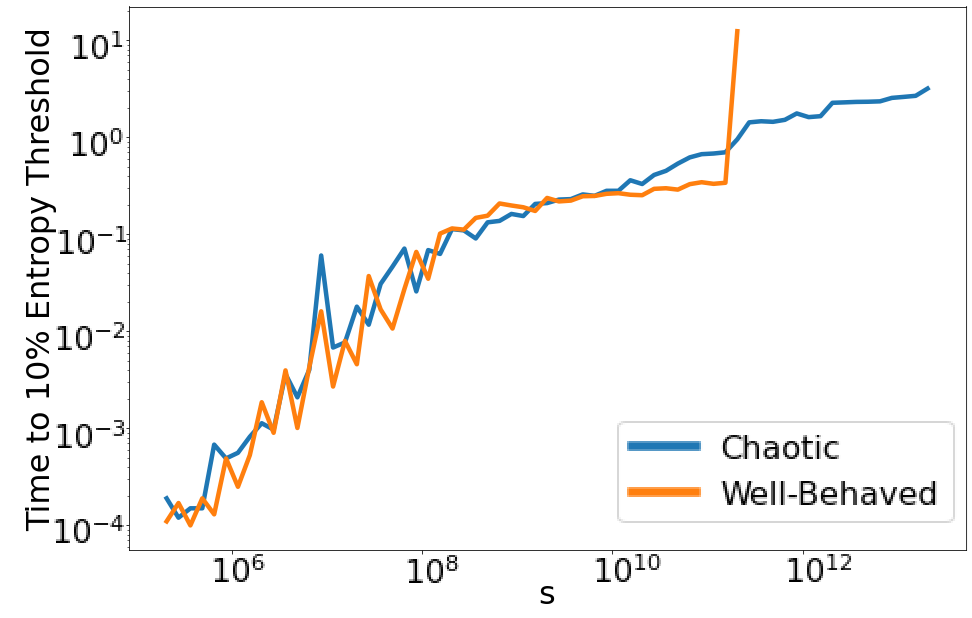}
    \caption{The timing of the first branch of the quantum algorithm's results, signaled by an increase to $10\%$ of the maximum entropy, as a function of $s$; at low $s$, the integrable (orange) and chaotic (blue) Lorenz systems scale similarly, but past a threshhold at $s\sim 10^{11}$, the integrable solutions converge and no longer branch; this phenomenon does not appear for the chaotic system.}
    \label{fig:EntropyScaling}
\end{figure}

Beyond the point of divergence, the quantum algorithm is no longer reliable, so if we desire reliable results for a set runtime $T$, we must find some method to delay the branching point. The most obvious method is to increase the stochastic sampling rate $s$; as $s$ increases, the algorithm should remain reliable longer. However, because the algorithm requires $MTs$ measurements, an incentive exists to minimize $s$ for a given runtime $T$. Figure \ref{fig:EntropyScaling} shows the time of divergence as a function of $s$. For the well-behaved system, an asymptote exists beyond which entropy never surpasses $10\%$ of its maximum value, because all solutions reach the fixed point. In other words, for the well behaved system, there exists an upper bound for $s$ beyond which error stays below a threshold for all time.  In comparison, for the chaotic system, error always exceeds a given threshold after some finite time. For the algorithm to be efficient, the dynamical system must have the property that $s$, the requisite stochastic sampling rate, increases slowly with $T$, the targeted runtime.

\section{Summary}

We have developed a method of mapping any real, polynomial dynamical system to a form which could be implemented on future fault-tolerant quantum hardware. This mapping results in a cubic, norm-preserving form which can be stored in a quantum state and evolves via a nonlinear Hamiltonian. We defined the nonlinear Hamiltonian as the sum of constant Hamiltonians, weighted by quadratic observables, and we approximate the nonlinear Hamiltonian as being constant over a small time period, allowing for the use of Hamiltonian simulation methods. By applying the algorithm to the logistic and Lorenz systems, we demonstrated that this map recovers the classical dynamics and can detect the onset of chaos. In the non-chaotic regime of the Lorenz system, a minimum value of the stochastic sampling rate was found to ensure that the simulated and deterministic solutions never diverge, but this cutoff was not observed in the chaotic system.

Of future interest are systems which are neither convergent nor chaotic. Under logistic dynamics and the well-behaved Lorenz system, early convergence to equilibrium prevents a deeper study of the algorithm's long-term error. Meanwhile, under chaotic dynamics, even classical algorithms are expected to diverge quickly. More study is required between these two extremes to determine long-term behavior for non-trivial, non-chaotic dynamics.

Whether this algorithm can be implemented efficiently on quantum hardware remains an open question. For systems which can only be realized with a large number of observables, or which include dense sub-Hamiltonians, it is unlikely that an algorithm of this form could be efficient. Variants of the algorithm presented here may efficiently solve classes of problems not covered by other existing quantum algorithms. However, more research is needed to further investigate the efficiency of a measurement-based approach. 

\textbf{Code Avalability:} The code developed for this paper is available via \cite{bib:GitLabRepo}.

\textbf{Competing Interests:} The authors declare none.

\section*{Acknowledgement}

This work was supported by the U.S. Department of Energy under Grant No. DE-SC0020393.

\appendix

\section{}\label{appA}

\subsection{Derivation of Norm-Preserving Dynamics}

Following \citet{bib:AlexThesis}, Eq.~(\ref{eqn:NormPreservingDynamics}) may be derived from fundamental vector-calculus rules. Following the rule for the product of a scalar-valued function and vector-valued function, we find
\begin{equation}
    \begin{split}
        \frac{d}{dt}\Big(\frac{\boldsymbol{x}}{|\boldsymbol{x}|}\Big) & = \frac{1}{|\boldsymbol{x}|}\frac{d}{dt}(\boldsymbol{x})+\boldsymbol{x}\frac{d}{dt}\Big(\frac{1}{|\boldsymbol{x}|}\Big) \\
         & = |\boldsymbol{x}|^{-1}\boldsymbol{G}(\boldsymbol{x})-|\boldsymbol{x}|^{-3}[\boldsymbol{x}\cdot\boldsymbol{G}(\boldsymbol{x})]\boldsymbol{x},
    \end{split}
    \label{eqn:ProductRule1}
\end{equation}
and using Eq.~(\ref{eqn:HomogeneousDynamicRatio}), that the system is homogeneous, we can pull $|\boldsymbol{x}|$ to the front:
\begin{equation}
    \frac{d}{dt}\Big(\frac{\boldsymbol{x}}{|\boldsymbol{x}|}\Big) = |\boldsymbol{x}|^{q-1}\Big(\boldsymbol{G}(\hat{\boldsymbol{x}})-[\hat{\boldsymbol{x}}\cdot\boldsymbol{G}(\hat{\boldsymbol{x}})]\hat{\boldsymbol{x}}\Big).
    \label{eqn:ProductRule2}
\end{equation}
Inserting a factor of $|\hat{\boldsymbol{x}}|^2$ to the first term to yield Eq.~(\ref{eqn:NormPreservingDynamics}) does not affect the system's dynamics because by definition, $|\hat{\boldsymbol{x}}|$ is always one. The addition of that factor ensures that the final expression is a homogeneous polynomial.

\subsection{Derivation of Anti-Symmetric Forms}

The derivation of Eq.~(\ref{eqn:FAntiSymmetry}) requires manipulation of the norm-preserving tensor system. For a tensor system, such as in Eq.~(\ref{eqn:TensorFormNP}), to be norm-preserving requires that
\begin{equation}
    \hat{\boldsymbol{x}}\cdot\frac{d}{dt'}\hat{\boldsymbol{x}} = \mathsfbi{F}x^{\otimes (q+3)} = 0,
    \label{eqn:XXdotOrthogonal}
\end{equation} 
containing an implicit sum of products in the entries of $\boldsymbol{x}$, and by the commutativity of multiplication, a given set of factors in $\boldsymbol{x}$ may correspond to more than one entry in $\mathsfbi{F}$ \citep{bib:AlexThesis}. However, because the entries in $\boldsymbol{x}$ are independent of each other, the only way for Eq.~(\ref{eqn:XXdotOrthogonal}) to be identically zero is for entries to be cancel. Thus, for a multi-index $\alpha$ of $q+3$ indices,
\begin{equation}
    \sum_{P\in S_{q+3}} \mathsfbi{F}_{P\alpha} = 0,
    \label{eqn:SumOfPermutations}
\end{equation}
where $S_{q+3}$ is the set of permutations on $q+3$ elements \citep{bib:AlexThesis}.

Because the dynamical system must eventually be mapped to a Hamiltonian, it is necessary to enforce a certain symmetry on the tensor form. Hamiltonians are Hermitian, but because Schr\"odinger's equation [Eq.~(\ref{eqn:SchrodingerEquation})] includes an extra imaginary factor, the matrices encoded in the tensor system must be anti-Hermitian. Hence, the tensor system must be antisymmetric in its first two indices. For a given, norm-preserving dynamical system, the form of the tensor $\mathsfbi{F}$ is not unique, and thus we find a method of converting a general $\mathsfbi{F}$ to this antisymmetric form. To enforce antisymmetry onto $\mathsfbi{F}$ without changing the system's dynamics \citep{bib:AlexThesis} takes

\begin{equation}
    \begin{split}
        \tilde{\mathsfbi{F}}_\alpha& \equiv \frac{1}{q+3}\Big(2(q+2)\mathsfbi{F}_\alpha-(q+1)\mathsfbi{F}_\alpha+\sum_{i=2}^{q+3}\mathsfbi{F}_{P_{1i}\alpha}-\sum_{i=2}^{q+3}\mathsfbi{F}_{P_{1i}\alpha}\Big) \\
         & \equiv \frac{1}{q+3}\Bigg[\sum_{i=2}^{q+3}\Big(\mathsfbi{F}_{\alpha}-\mathsfbi{F}_{P_{1i}\alpha}\Big)+\sum_{i=2}^{q+3}\Big(\mathsfbi{F}_{\alpha}+\mathsfbi{F}_{P_{1i}\alpha}\Big)-(q+1)\mathsfbi{F}_\alpha\Bigg].
    \end{split}
    \label{eqn:AntiSym1}
\end{equation}

Again, by the commutativity of multiplication, the indices of $\mathsfbi{F}$, with the exception of the first index, can be reordered without affecting the system's dynamics. Thus, Eq.~(\ref{eqn:AntiSym1}) is equivalent to
the sum of $\frac{1}{q+3}\sum_{i=2}^{q+3}(\mathsfbi{F}_{P_{2i}\alpha}-\mathsfbi{F}_{P_{1i}P_{2i}\alpha})$ and $\frac{1}{q+3}[\mathsfbi{F}_\alpha+\sum_{i=2}^{q+3}\mathsfbi{F}_{P_{1i}\alpha}]$ \citep{bib:AlexThesis}.

The second term $\frac{1}{q+3}[\mathsfbi{F}_\alpha+\sum_{i=2}^{q+3}\mathsfbi{F}_{P_{1i}\alpha}]$ reduces to $\frac{1}{q+3}\mathsfbi{T}_\alpha\equiv\frac{1}{q+3}\sum_{i=1}^{q+3}\mathsfbi{F}_{P_{1i}\alpha}$. Notably, because the order of the final $q+2$ indices are arbitrary, we can replace $\mathsfbi{T}_{j\beta}$ with
\begin{equation}
    \tilde{\mathsfbi{T}}_{j\beta} \equiv \frac{1}{(q+2)!}\sum_{P\in S_{q+2}}\mathsfbi{T}_{jP\beta}
    \label{eqn:SymmetricT}
\end{equation}
for any multi-index $\beta$ of $q+2$ indices; the effect of Eq.~(\ref{eqn:SymmetricT}) is to enfore symmetry in $\mathsfbi{T}$ across the latter indices. We see then
\begin{equation*}
    \frac{1}{(q+2)!}\sum_{P'\in S_{q+2}}\mathsfbi{T}_{jP\beta} = \frac{1}{(q+2)!}\sum_{P''\in S_{q+3}} \mathsfbi{F}_{P''[j\beta]};
    \label{eqn:KillerT}
\end{equation*}
because the permutation $P'$ shuffles the latter indices and the $P_{1i}$ swaps the first index into any of the latter positions, all possible ordering of the multi-index $j\beta$ are reached. Thus, by Eq.~(\ref{eqn:SumOfPermutations}), the second term in the sum for $\tilde{\mathsfbi{F}}_\alpha$ must be zero, yielding Eq.~(\ref{eqn:FAntiSymmetry}):
\begin{equation*}
    \tilde{\mathsfbi{F}}_\alpha \equiv \frac{1}{q+3}\sum_{i=2}^{q+3}(\mathsfbi{F}_{P_{2i}\alpha}-\mathsfbi{F}_{P_{1i}P_{2i}\alpha}).
    \label{eqn:AntiSym3}
\end{equation*}
The two sides of Eq.~(\ref{eqn:FAntiSymmetry}) yield equivalent dynamics, but the right side is in a form which is antisymmetric in the first two indices.

\subsubsection{A Simple Example}

To demonstrate how Eq.~(\ref{eqn:FAntiSymmetry}) can be applied, let us first consider the following norm-preserving system:
\begin{equation}
    \begin{split}
        \frac{d}{dt'} \hat{x}_1 & = \hat{x}_2\hat{x}_3, \\
        \frac{d}{dt'} \hat{x}_2 & = -\hat{x}_1\hat{x}_3, \\
        \frac{d}{dt'} \hat{x}_3 & = 0. \\
    \end{split}
    \label{eqn:SimpleExample1a}
\end{equation}
The tensor system $\mathsfbi{F}_{123} = 1$, $\mathsfbi{F}_{213} = -\frac{1}{2}$, $\mathsfbi{F}_{231} = -\frac{1}{2}$ realizes the dynamics in Eq.~(\ref{eqn:SimpleExample1a}). Following through Eq.~(\ref{eqn:FAntiSymmetry}) yields the non-zero elements of an equivalent, antisymmetric tensor:
\begin{equation}
    \begin{split}
        \mathsfbi{A}_{123} & = \frac{1}{3}\Big[(\mathsfbi{F}_{123}-\mathsfbi{F}_{213})+(\mathsfbi{F}_{132}-\mathsfbi{F}_{231})\Big] = +\frac{2}{3}, \\
        \mathsfbi{A}_{132} & = \frac{1}{3}\Big[(\mathsfbi{F}_{132}-\mathsfbi{F}_{312})+(\mathsfbi{F}_{123}-\mathsfbi{F}_{321})\Big] = +\frac{1}{3}, \\
        \mathsfbi{A}_{213} & = \frac{1}{3}\Big[(\mathsfbi{F}_{213}-\mathsfbi{F}_{123})+(\mathsfbi{F}_{231}-\mathsfbi{F}_{132})\Big] = -\frac{2}{3}, \\
        \mathsfbi{A}_{231} & = \frac{1}{3}\Big[(\mathsfbi{F}_{231}-\mathsfbi{F}_{321})+(\mathsfbi{F}_{213}-\mathsfbi{F}_{312})\Big] = -\frac{1}{3}, \\
        \mathsfbi{A}_{312} & = \frac{1}{3}\Big[(\mathsfbi{F}_{312}-\mathsfbi{F}_{132})+(\mathsfbi{F}_{321}-\mathsfbi{F}_{123})\Big] = -\frac{1}{3}, \\
        \mathsfbi{A}_{321} & = \frac{1}{3}\Big[(\mathsfbi{F}_{321}-\mathsfbi{F}_{231})+(\mathsfbi{F}_{312}-\mathsfbi{F}_{213})\Big] = +\frac{1}{3}. \\
    \end{split}
    \label{eqn:SimpleExample1c}
\end{equation}
Notably, this tensor has more non-zero elements than necessary. An alternative with the minimum number of non-zero elements is $\mathsfbi{A}_{123} = +1$, $\mathsfbi{A}_{213} = -1$. Tensor forms can be made more sparse by restricting the order of latter indices, namely the second through last indices of the input tensor $\mathsfbi{F}$ and the third through last of $\mathsfbi{A}$. Although not optimal, the tensor in Eq.~(\ref{eqn:SimpleExample1c}) can be mapped to a Hamiltonian.

\subsection{Derivation of Degree-Reduced System}

Eq.~(\ref{eqn:MDegreeReduction3}) is one implementation of a system which satisfies fundamental calculus rules for a product, because \citep{bib:AlexThesis}:
\begin{equation}
    \begin{split}
        \frac{d}{dt'}\hat{\boldsymbol{y}}_\alpha & = \frac{d}{dt'} \prod_{i=1}^{(q+1)/2}\hat{\boldsymbol{x}}_{\alpha_i} \\
         & = \sum_{i=1}^{(q+1)/2} \Big(\prod_{j\neq i}\hat{\boldsymbol{x}}_{\alpha_j}\Big)\frac{d}{dt'}\hat{\boldsymbol{x}}_{\alpha_i} \\
         & = \sum_{i=1}^{(q+1)/2} \Big(\prod_{j\neq i}\hat{\boldsymbol{x}}_{\alpha_j}\Big)\Big(\sum_{\beta}\mathsfbi{A}_{\alpha_i\beta}\prod_{k=1}^{q+2}\hat{x}_{\beta_k}\Big). \\
    \end{split}
    \label{eqn:DegreeReductionDerivation}
\end{equation}

\subsubsection{A Simple Example}

To demonstrate the use of Eq.~(\ref{eqn:MDegreeReduction3}), let us consider the following antisymmetric, norm-preserving tensor system:
\begin{equation}
    \begin{split}
        \mathsfbi{A}_{121111} & = -1, \\
        \mathsfbi{A}_{211111} & = +1.
    \end{split}
    \label{eqn:SimpleExample2a}
\end{equation}
For a rank-$6$ $\mathsfbi{A}$, the corresponding $\mathsfbi{M}$ must have $4$ multi-indices of $2$ indices each. In the case of only two variables, as in Eq.~(\ref{eqn:SimpleExample2a}), this corresponds to a $4\times 4\times 4\times 4$ $\mathsfbi{M}$ tensor.

For a sparse $\mathsfbi{A}$, as in Eq.~(\ref{eqn:SimpleExample2a}), Eq.~(\ref{eqn:MDegreeReduction3}) determines which elements of $\mathsfbi{M}$ are non-zero. First, the indices of $\mathsfbi{A}$ are split into four multi-indices, with the first and second indices are part of $\alpha$ and $\beta$, respectively, the third and fourth indices being $\nu$, and the fifth and sixth indices being $\eta$. The full set of multi-indices of $M$ are determined then by filling out $\alpha$ and $\beta$ with another index, to create multi-indices of $2$ indices. By the product of Kronecker deltas, these added indices must be identical and added in the same position of the multi-index. For example, the element $\mathsfbi{A}_{(1)(2)(11)(11)}$ would contribute to the elements $\mathsfbi{M}_{(11)(21)(11)(11)}$ and $\mathsfbi{M}_{(21)(22)(11)(11)}$, but not $\mathsfbi{M}_{(11)(22)(11)(11)}$ or $\mathsfbi{M}_{(12)(12)(11)(11)}$. Thus, each of the elements in Eq.~(\ref{eqn:SimpleExample2a}) corresponds to four elements in the final $\mathsfbi{M}$, corresponding to adding the index $1$ ($2$) to the first (second) position in $\alpha$ and $\beta$; these four elements may not be unique. The eight non-zero elements of $\mathsfbi{M}$ in the system are then
\begin{equation}
    \begin{split}
        \mathsfbi{M}_{(11)(12)\nu\eta} = \mathsfbi{M}_{(11)(21)\nu\eta} = \mathsfbi{M}_{(21)(22)\nu\eta} = \mathsfbi{M}_{(12)(22)\nu\eta} & = \mathsfbi{A}_{(1)(2)\nu\eta} = -1, \\
        \mathsfbi{M}_{(12)(11)\nu\eta} = \mathsfbi{M}_{(21)(11)\nu\eta} = \mathsfbi{M}_{(22)(21)\nu\eta} = \mathsfbi{M}_{(22)(12)\nu\eta} & = \mathsfbi{A}_{(2)(1)\nu\eta} = +1,
    \end{split}
    \label{eqn:SimpleExample2b}
\end{equation}
where $\nu=\eta=(11)$.

\bibliographystyle{jpp}

\bibliography{jpp-instructions}

\end{document}